# Motivation and Development of a Compact Superconducting Accelerator for X-ray Medical Device Sterilization

**Thomas K. Kroc**
Fermi National Accelerator Laboratory
PO Box 500, Batavia, IL 60510
kroc@fnal.gov

**ABSTRACT**

Fermilab is developing a compact superconducting (CSRF) accelerator as a source of a high-power, high-energy electron beam to produce an x-ray beam comparable to $\geq 2$ MCi of cobalt-60. As part of this development, we are presently assembling a 20 kW, 1.6 MeV prototype that incorporates the four enabling technologies described below. These technologies support a heat budget that is within the capacity of commercially available cryocoolers and eliminate the need for liquid cryogens. The use of superconducting technologies promises efficiencies, wall-plug to beam, of 80% or greater. A summary of the design and progress to date will be given. In addition, we will present a review of developments within the medical device sterilization industry that are motivating the development of this technology.

## 1. INTRODUCTION

Approximately 50% of single use medical devices are sterilized with gamma rays from the decay of Co-60. Worldwide, there is over 400 MCi of cobalt installed in facilities in over 40 countries. However, there is permitted capacity of around 600 MCi for those facilities [1]. The medical device industry is growing at ~7% per annum. This growth projection does not include the recent impact of COVID-19. The sterilization market is very tight with little or no capacity to deal with interruptions or unexpected shutdowns of facilities and sterilization capacity shortages are looming. The production of cobalt-60 is presently behind market demand by 5% [1] and given the cycle times of cobalt production, it may take a few years for production to catch up. Additionally, there are radioisotope security concerns that lead some to press for a reduction in our dependence on radioisotopes.

As can be seen in Fig. 1, the penetration characteristics of x-rays produced by a 7.5 MeV electron beam is very similar to the penetration of the gamma rays from cobalt-60. This means that x-rays can be a direct replacement for gamma sterilization of medical devices. The Bio-Process Systems Alliance (BPSA), an organization representing manufacturers of single use systems for the production of vaccines and pharmaceuticals, is aggressively pursuing the use of x-ray sterilization of their products by the end of 2022.

Superconducting accelerators used to produce x-rays offer high-power and efficient sources of ionizing radiation to effectively sterilize all products that are presently sterilized using gamma rays from cobalt-60.

Their efficiency comes from two areas: 1) the efficiency of the accelerating system itself, and 2) the efficient utilization of energy to produce the RF energy needed to drive the accelerator.

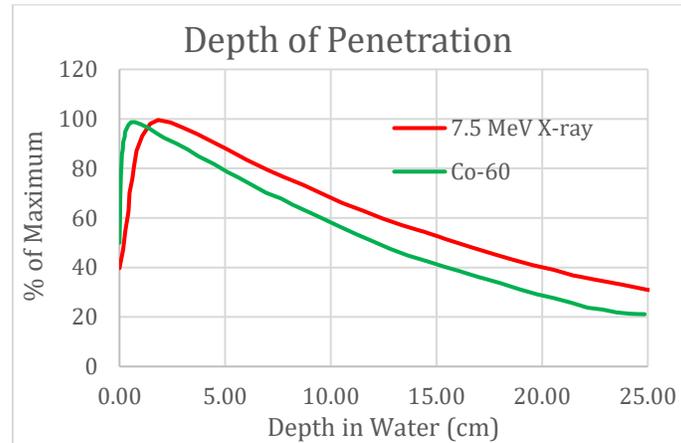

**Figure 1. Comparison of the depth of penetration of photons from cobalt-60 decay and Bremsstrahlung x-rays from a 7.5 MeV electron beam in water.**

One megacurie (MCi) of cobalt-60 produces about 15 W of power in the gamma rays that are released. Due to the inefficiency of the Bremsstrahlung process, it requires about 120 kW of electron beam power onto a high-Z target to produce the same amount of power in the resulting x-rays. Conventional linacs have not yet approached the power necessary to compete with multi-megacurie cobalt facilities.

Normal conducting accelerators (which operate at room temperature) produce inductive currents in the conducting, inner surface of the accelerating cavities. These currents result in the generation of heat and the dissipation of energy due to the resistance of the cavity material (typically copper). The amount of power lost can be greater than the power of the beam of the accelerated particles. In a superconducting system, the resistance of the cavity surface is negligible resulting in almost all the supplied power being utilized in the power of the resulting electron beam.

## 2. REQUIREMENTS OF NEW ACCELERATOR DESIGNS

Economic factors will be the deciding factors for the adoption of new technologies for sterilization systems. Efficiency plays a very large role in the economics of a system and is investigated in depth in the next section. There are, however, other parameters that play into a prospective user's economic analysis. Some of these parameters are: robustness and ease of use; product throughput (related to the Linac power); optimized conveyor systems; Beam energy penetration (related to the Linac energy) and the efficiency of e-beam to X ray conversion (Bremsstrahlung efficiency).

Accelerator systems have more active components than cobalt arrays and therefore need to be highly reliable so that maintenance can be carried out on a predicable schedule rather than having to respond to breakdowns. They should not have to rely on specialized technical staff to operate or maintain.

While certain models of accelerators have sufficient power (~350 kW) to compete with many cobalt facilities, current linac systems (~ 100 – 150 kW) need higher power to better mesh with existing supply chains. Further optimization of conveyor systems for X-ray facilities appears to be possible to ensure that maximum usage of the X-ray photons is achieved.

The efficiency of the Bremsstrahlung process (~12% at 7 MeV) is constrained by fundamental physics. It may be useful to consider increasing the allowable maximum energy of x-rays in order to increase the Bremsstrahlung efficiency. This must be balanced against radiation safety requirements.

## 3. THE CASE FOR SUPERCONDUCTIVITY

An economic analysis of the cost of gamma and x-ray facilities suggests that long term costs for an x-ray facility are less than for gamma when the effective capacity is greater than 1–2 MCi. [3]. Despite this, any economic advantage may not be significant enough to overcome other expenses of revalidating existing medical devices. This results in ambivalence in modality from a cost perspective for sterilizing new medical devices and provides little incentive to revalidate existing devices in a new modality. Improving the electrical efficiency of accelerator systems would improve this comparison and there is considerable room for improvement in this area.

An analysis of the factors that drive the efficiency of accelerator systems has been conducted [2]. Figure 2 compares the relationship of the overall accelerator system efficiency with the RF source efficiency for three accelerator systems, one at two different duty factors, to an SRF system. The parameters used to calculate the curves are included in Table 1. Additional parameters include, 5% was assumed for RF control and 10% for line losses. Fifteen percent of the RF power was added for RF system cooling and 15% of the Ohmic losses were added for linac cooling. The lines in Figure 2 show the maximum theoretical efficiency. The advantage of superconducting systems (SRF) is clearly shown. While there are accelerator systems that currently can provide beam powers that are competitive with megacurie cobalt systems, their efficiency is significantly poorer than SRF. For an example of a linac system, a L-Band linac operating at room temperature was used as the necessary parameters for the calculation were readily available. The performance of newer linac systems is not expected to be significantly different. This accelerator, when operating at 5% duty factor, shows a performance that approaches that of the SRF system. However, it requires an RF source with a peak power of 5.5 MW. This peak power requirement can be reduced by increasing the duty factor at the cost of system efficiency.

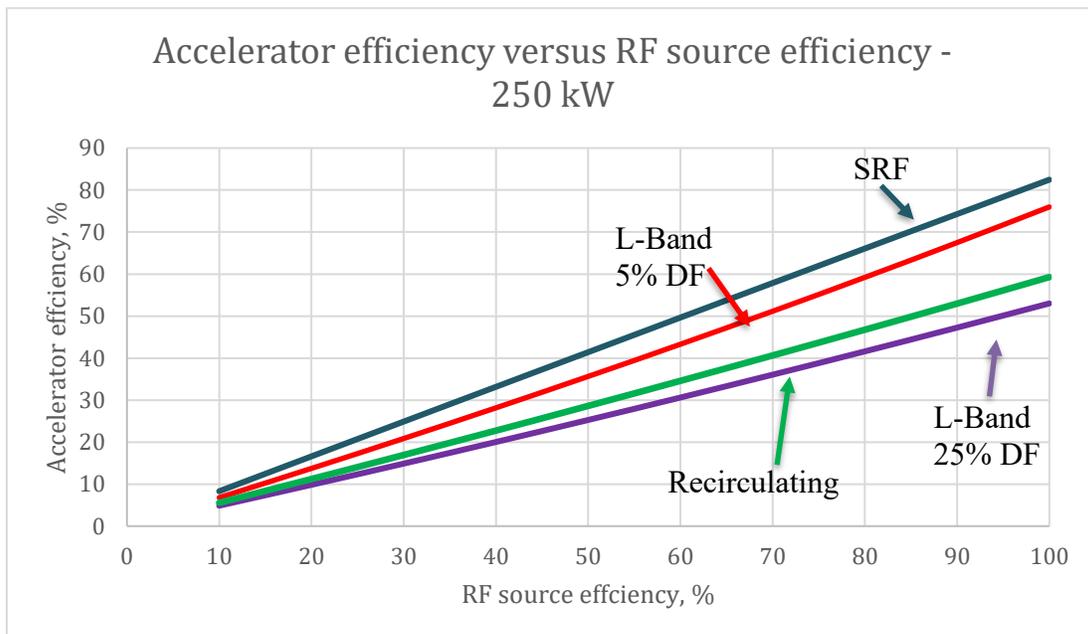

**Figure 2. Comparison of accelerator system efficiencies for superconducting accelerator, room temperature linac with two separate duty factors (DF), and a recirculating VHF accelerator.**

| | | SRF Nb₃Sn (DC) | L-Band Linac | L-Band Linac | Recirculating VHF (DC) |
|---|---|---|---|---|---|
| Frequency | MHz | 650 | 1300 | 1300 | 107.5 |
| Length | m | 1.6 | 3.25 | 3.25 | 2 |
| Energy | MeV | 10 | 10 | 10 | 10 |
| Average beam power | kW | 250 | 250 | 250 | 250 |
| Duty Factor | % | 100 | 5 | 25 | 100 |
| Pulsed beam power | kW | 250 | 5000 | 1000 | 250 |
| Power for refrigeration | kW | 40 | N/A | N/A | N/A |
| Pulsed Ohmic loss in the linac | kW | N/A | 540 | 540 | 86 (DC) |
| Pulsed RF power | kW | N/A | 5540 | 1540 | N/A |
| Average RF power | kW | 250 | 277 | 385 | 336 |
| Average Ohmic losses | kW | N/A | 27 | 135 | 86 |
| Beam current | mA | 25 | 500 | 100 | 25 |

Table 1. Parameters used in calculation of accelerator system efficiency.

Figure 3 shows the impact of the efficiency of two common RF sources on the total system efficiency. Solid state RF sources are becoming more common; however, they are only about 45% efficient. Magnetron RF sources are inherently much more efficient, at approximately 85%. The impact of this difference is greater than may be initially apparent when just comparing 45% with 85%. The beam power required to produce x-rays equivalent to 1 megacurie of cobalt is approximately 120 kW. At 45% efficiency, this requires 267 kW of input power. This means 147 kW of power is wasted and must be removed by cooling systems. At 85% efficiency, the total power required is 141 kW which means only 21 kW of waste power must be removed. This is a reduction of a factor of 6.7. Note that this does not include the inefficiency of the Bremsstrahlung converter which is the same for all systems.

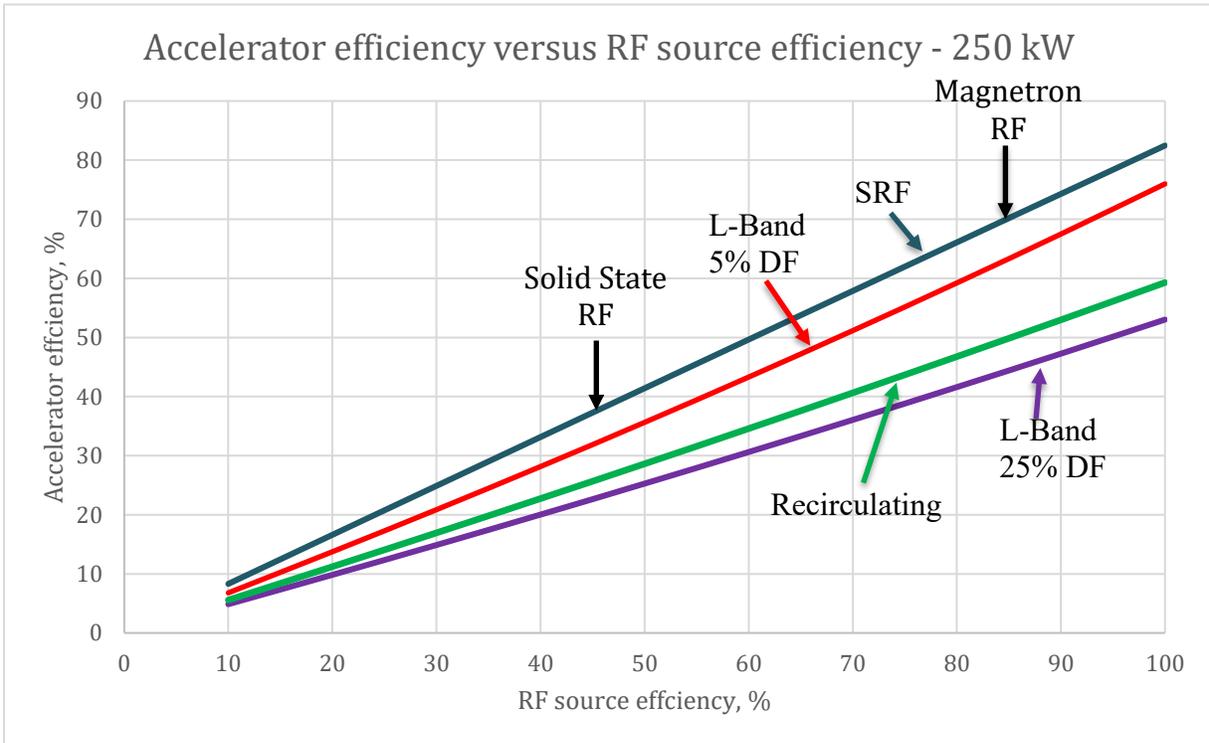

Figure 3. The impact of two common RF sources on total system efficiency.

## 4. FERMILAB DEVELOPMENT

The efficiency, size, weight, and power, of superconducting accelerators comes from the integration of 4 enabling technologies. Three of these minimize the generation or introduction of heat into the system and one removes that heat from the system without the need for liquid cryogens.

Minimize heat generation

- Integrated electron source
- Low heat-loss RF coupler
- $Nb_3Sn$ coated cavities

Efficiently remove heat

- Conduction cooling & commercial cryocoolers

The integrated electron source does not use a Low Energy Beam Transport (LEBT) line. Rather the thermionic cathode is positioned in a small opening in the superconducting cavity. Despite the high temperature of the cathode (~ 1050 C), its small size and careful design limit the amount of heat introduced into the system to 0.3 W (for our 20 kW prototype noted further below).

The low heat-loss coupler is also a connection from the cryogenic portion of the superconducting cavity to the ambient environment. Thin walls and an intermediate temperature intercept limit the conducted heat to 1 W. Additionally, the use of solid copper for the electromagnet structures eliminates the risk of flaking from electroplated elements. These flakes can spoil the conductance of the coupler.

The most critical aspect of the superconducting design is the use of a $Nb_3Sn$ coating on the inner surface of the accelerating cavity. Superconducting cavities have frequency dependent currents, which increase with the square of the frequency, that are induced in the walls of the cavity. Our choice of 650 MHz results in a heat generation of about 1.5 W. The coating doubles the critical temperature of the surface which allows operation at ~4K instead of 2K. Operating at 4K allows the use of cryocoolers to remove the heat generated by these currents. A critical measure of the performance of superconducting cavities is the quality factor $Q_0$ which is the ratio of the initial stored energy to the energy lost in each oscillation of the RF. We have demonstrated cavities with a $Q_0$ of greater than $1 \times 10^{10}$ with an average electric field gradient of about 7 MV/m. This is a very conservative gradient compared to accelerators used in discovery science and is a key element in the design to provide robustness for use in industrial environments.

Commercial cryocoolers are now able to dissipate 2–2.5 W of heat at 4K, which is enabled by the $Nb_3Sn$ coating as noted above. These coolers operate without liquid cryogens and the large infrastructure needed to produce and maintain the cryogens. To efficiently move the heat from its origin, a method, called conduction cooling, has been developed [5–9] to conductively transfer the heat to the cryocoolers. This uses high purity aluminum and required a thorough investigation of proper design, sizing, and fastening to attain a fraction of a degree temperature change across the transfer path.

For the example systems used to calculate the data for Figure 3, 40 kW is required to power the cryocoolers to remove the <4W of heat from the superconducting system. The normal conducting systems generate 540 kW of heat through Ohmic losses that must be removed by conventional cooling systems.

To validate the compact SRF concept, we are in the process of assembling a 1.6 MeV, 20 kW prototype that incorporates all of the enabling technologies noted above. As shown in Figure 4, the cryostat is about 1 meter long and 1 meter in diameter. Two cryocoolers (one shown) will provide the necessary cooling capacity. It has one full-length accelerating cavity and one partial cavity. The partial cavity matches the beam emitted from the electron source to the full cavity. We refer to this combination as a 1.5 cell cavity. This is the minimal system to illustrate the successful integration. Assembly should be completed in late 2023. It is possible that this device itself may have commercial applications.

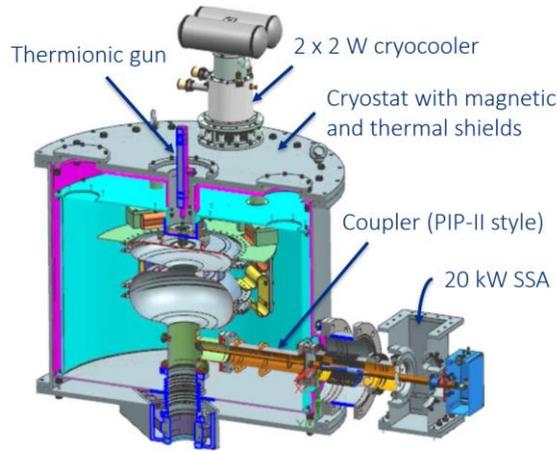

**Figure 4. Diagram of 1.6 MeV, 20 kW prototype with 1.5 cell Nb$_3$Sn accelerating cavity, integrated electron gun, and low heat-loss RF coupler.**

## 5. EFFICIENCY VS FLEXIBILITY

With the successful demonstration of the 20 kW prototype, the design will be ready for the production of a commercial scale system (200 kW) that can produce the equivalent of 2 MCi of cobalt-60 or more. We have developed a digital design of such a system. A preliminary illustration of the system is shown in Figure 5. While an actual system will not be quite as compact as shown here, it will still be more compact than existing linac systems of comparable power. This system utilizes 5.5 cells to achieve an electron beam energy of 7.5 MeV while still using the conservative gradient noted previously. The 5.5 cells generate 6.4 watts of heat compared to the 1.5 watts in the 1.5 cell cavity. This system will require 4 cryocoolers, so the gold structure in the figure will require two racks instead of one. Two 100 kW magnetrons would be housed in the light blue rack. Controls and vacuum supplies would occupy the dark rack. This system would supply the electron beam to a scanning horn with a Bremsstrahlung converter as is presently used in present x-ray sterilization applications.

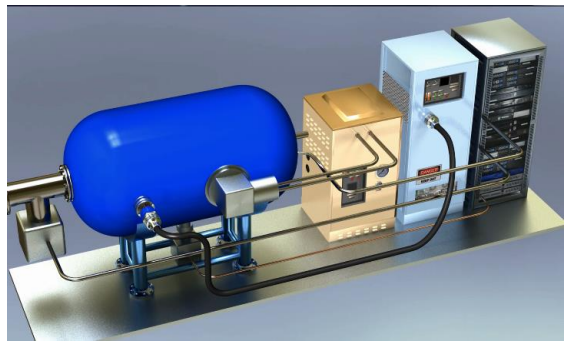

**Figure 5. Conceptual illustration of 200 kW compact SRF accelerator system including cryocooler compressor, control racks, and magnetron RF supply.**

## 6. RF NEEDS

In order to fully realize the efficiency potential of the superconducting system, efficient RF power supplies are required as noted in Section 3 above. Solid state power supplies are becoming a favorite of the market. However, the electrical efficiency of these supplies is less than 50%. Magnetrons are quite simple, inexpensive and have electrical efficiencies of 80% or more. However, the performance requirements for currently available commercial magnetrons may not be suitable for these high-performance applications, particularly their expected lifetime, which can be quite variable at present. To that end, we have identified a development plan to improve the performance of magnetrons while still being able reap the economy of these devices.

Magnetron tubes are inexpensive, so the cost of replacement is not a concern in present commercial applications such as the process heating industry (drying wallboard and lumber). For an industry such as medical device sterilization, the concern would be process interruption on a frequent and irregular basis. Longer lifetime and a more dependable mean-time-to-failure may be necessary for adoption of this power source. The main reasons for the present state of tube lifetime are:

- Anode sputtering of the cathode material
- Cathode bombardment by backward electrons.

General measures which may be taken to address these issues are:

- Active vacuum pumping of the magnetron
- Electron dynamics optimization.

Attention to electron dynamics may also improve the magnetron efficiency. Recent investigations show [10, 11] that magnetron efficiency may be improved together with lifetime extension by operating in a sub-critical regime.

While our present efforts are focused on completing the integration of a complete CSRF system, we have outlined a program to address the issues of magnetron lifetime and efficiency.

The program:

- Use a modern 3D simulation code to understand in detail the beam dynamics of a magnetron.
- Benchmark the code with experimental measurements. This improved and benchmarked code will strengthen the RF industry allowing better designs of the magnetron for different applications – scientific, industrial, civil, and military.
- Finally, it would be possible to optimize the magnetron design to improve its longevity and efficiency and optimize various operation regimes. Different options could be explored, like 2D harmonic cavities, different types of cathodes including the newly developed Nanocomposite Scandate Tungsten cathodes [12].

The goal would be to achieve an efficiency of more than 85% with reliable tube lifetime of ~50,000-80,000 hours.

# 7. CONCLUSION

The need for new sources of ionizing radiation for medical device sterilization is acute. New technology is arriving to provide higher-power, more economical x-ray sources. These x-ray sources can provide direct alternatives to gamma ray sources. The compact superconducting accelerator system described here utilizes four enabling technologies to provide a new generation of linac sources to meet this challenge.

**ACKNOWLEDGMENTS**

Thanks to Vyacheslav Yakovlev for calculating the efficiency graphs of additional accelerator examples.

This manuscript has been authored by Fermi Research Alliance, LLC under Contract No. DE-AC02-07CH11359 with the U.S. Department of Energy, Office of Science, Office of High Energy Physics.